\documentstyle[12pt]{article}
\newcommand{\be}{\begin{equation}}
\newcommand{\ee}{\end{equation}}

\begin{document}

\vspace*{4cm}

\begin{center}
\large{\bf CONSTITUENT STRING MODEL FOR
HYBRID MESONIC EXCITATIONS }\\

\vspace{1cm}

{\bf Yu.S.Kalashnikova\footnote{E-mail: yulia@vxitep.itep.ru},
Yu.B.Yufryakov\footnote{E-mail: yufryakov@vxitep.itep.ru}} \\

\vspace{0.5cm}

{\small Institute of Theoretical and Experimental Physics\\
117259, Russia, Moscow, B.Cheremushkinskaya 25.}

\end{center}
\vspace{1cm}
\begin{abstract}

The model for hybrid excitations of the QCD string with
quarks is presented starting from the perturbation theory
in the nonperturbative background. The propagation of a system
containing $q\bar q$--pair and  gluon is  considered. The
simplified version of the Hamiltonian, including both long--range
nonperturbative interaction and Coulomb force, is derived. The masses
of the lowest
$q\bar{q} g$ hybrids are evaluated, and numerical results for the
spectra are listed.
 \end{abstract}

\section{Introduction}

One of the most important features of the QCD  Lagrangian is   the
presence of gluonic degrees of freedom which should exhibit
themselves at the constituent level, namely in the form of glueballs
and hybrids, i.e. non--$q\bar q$ exotics. Unfortunately, the current
experimental situation  is too complicated to give us unambigious
proof that such states exist; some candidates for non--$q\bar q$
exotics appear and disappear from time to time, changing their masses
and quantum numbers (for the up--to--date review see [1]).
Nevertheless, there is no doubts that standard $q\bar q$ nonets are
overpopulated, but the question about the nature of the "extra"
states is still open.  On the other hand, there is no substantial
progress in the description of  the strong coupling nonperturbative
regime in QCD. There are some QCD -- inspired theoretical approaches,
but none of them are able to provide reliable enough predictions for
masses and decay rates of hybrid mesons.

 The QCD sum rules estimations for exotic hybrids are rather
 unstable: first results [2] predicted for the $1^{-+}$  light hybrid
 the mass  1.2-1.7 GeV, while more  recent calculations give 2.1  GeV
 [3]  and 2.5 GeV [4].

 In the bag model the gluons are automatically transverse, and the
 lowest electric gluon (with  $J^P=1^+$) in the  spherical cavity is
 much lighter than the lowest magnetic one (with $J^P=1^{-}$). The
 lowest hybrids  with light quarks have the mass about 1.5 GeV [5,6].
 The masses of hybrids with heavy quarks were estimated in [7] taking
 into account the bag deformation, with the results 3.9 GeV
 for $c\bar c$ hybrid and 10.5 GeV for $ b\bar b$ one.

 Constituent gluon model was introduced in [8,9]. In this model the
 linear potential is introduced {\it ad hoc},  in analogy with the
 charmonium system. It was proposed also that the gluon orbital
 momentum is diagonal. As a consequence, in such model the lowest
 states with the  $S$--wave gluon have non--exotic quantum numbers.
 The mass predictions of constituent model are
  1.3-1.8 GeV for the
 lowest non--exotic hybrids.

Flux--tube model [10] predicts degenerate (up to spin corrections)
lightest hybrid states at 1.8-1.9 GeV. In this model phonon--type
excitations of the string connecting quark and antiquark are
interpreted as hybrids. First calculations [10] assumed small
oscillation approximation, and recent results [11] demonstrate that
this  approximation might be inadequate. Improved version [11] is
given in the framework of "one--bead" flux--tube model.

Here we present the studies of the  $q\bar q g$ system in the
framework of Vacuum Background Correlator method [12]. The main
assumption is that nonperturbative background fields $\{B_{\mu}\}$
exist in the QCD vacuum, which ensure the area law asymptotic for the
Wilson loop along the closed contour $C$,
$$
<W(C)>_B\to N_c exp(-\sigma S),
$$
with $S=S_{min}$ being the minimal surface inside the contour $C$. It
was confirmed by cluster expansion method [12] that if one assumes
the existence of finite correlation length for  the background, the
asymptotical behaviour of the Wilson loop average is compatible with
the area law. The deviation from the area law at large distances with
$S=S_{min}$ are caused by the perturbations over the background. The
"minimal" area law gives the string--type interaction in the $q\bar
q$ system, while the perturbative  fields are responsible for the
string vibrations [13].

\section{Green function for the $ q\bar q g $ system in the Vacuum
Background Correlator method}

The  constituent gluon in the Vacuum Background Correlator method is
introduced as a gluon propagating in the nonperturbative background
field [14,15]. Following [15], we split the gluonic field $A_{\mu}$
into the background field $B_{\mu}$ and the perturbation $a_{\mu}$
over the background.

  \be
  A_{\mu}=B_{\mu}+a_{\mu}
  \ee

  We ascribe the inhomogeneous part of gauge transformation to the
  field $B_{\mu}$,
  \be
  B_{\mu}\to U^+(B_{\mu}+\frac{i}{g}\partial_{\mu})U,~~
  a_{\mu}\to U^+a_{\mu}U,
  \ee
  so that the states involving the field $a_{\mu}$ may  be formed in
  the gauge--invariant manner. One--gluon hybrid is represented as
  \be
  \Psi(x_q,x_{\bar q}, x_g)=
  \bar{\psi}_{\alpha}(x_{\bar q})\Phi^{\alpha}_{\beta}(x_{\bar
  q},x_g) a^{\beta}_{\gamma}(x_g)\Phi^{\gamma}_{\delta}
  (x_g,x_q)\psi^{\delta}(x_q),
  \ee
  where $\alpha...\delta$ are the colour indices in the fundamental
  representation,
  $a^{\beta}_{\gamma}=a_a(\lambda_a)^{\beta}_{\gamma}$, and parallel
  transporters $\Phi$ contain only background field:
  \be
  \Phi^{\alpha}_{\beta}(x,y)=(P exp \int^x_y
  B_{\mu}dz_{\mu})^{\alpha}_{\beta}
  \ee

  The Green function for the $q\bar q g$ system is obtained by
  averaging the product $\Psi_{in}\Psi^+_{out}$ over the background
  field configurations:
  \be
  G
  (x_qx_{\bar q} x_g,
  y_qy_{\bar q} y_g)=
  <\Psi_{in}(
  y_q,y_{\bar q}, y_g)\Psi^+_{out}
  (x_qx_{\bar q} x_g)>_B.
  \ee

  The dynamics of the field $a_{\mu}$ is  defined, in accordance with
  the decomposition (1), by expanding the QCD Lagrangian up to the
  second order in the fields $a_{\mu}$ (in  the Euclidean
  space--time)
  \be
  L(a)=-\frac{1}{4}(F_{\mu\nu}(B))^2+ a_{\nu}D_{\mu}(B)F_{\mu\nu}(B)+
  \ee
  $$
  +\frac{1}{2}a_{\nu}
  (D_{\lambda}D_{\lambda}\delta_{\mu\nu}-D_{\mu}D_{\nu}-g
  F_{\mu\nu}(B))a_{\mu},
  $$
  $$
  D_{\lambda}^{ca}=\partial_{\lambda}\delta^{ca}+gf^{cba}B_{\lambda}^b,
  $$
  with the background gauge fixing term
  \be
  G^a=\partial_{\mu}a_{\mu}^a+gf^{abc}B_{\mu}^ba^c_{\mu}
  \ee

  In what follows we skip the issue of ghosts. The linear in the
  fields $a_{\mu}$ part of the Lagrangian (6) disappears if the field
  $B_{\mu}$ satisfies the classical equation of motion
  $D_{\mu}F_{\mu\nu}=0$.
 To be on the safe side one is to assume that the background is the
  classical one, or at least, that the transition vertex generated
  by this term is small.

  The Green function for the field $a_{\mu}$ propagating in the given
  background $B_{\mu}$ may be identified in the background gauge as
  \be
  G^{-1}_{\mu\nu}=D^2\delta_{\mu\nu}-D_{\nu}D_{\mu}-gF_{\mu\nu}+
  \frac{1}{\xi}D_{\mu}D_{\nu}=
  M_{\mu\nu}+\frac{1}{\xi}D_{\mu}D_{\nu} .
  \ee
  If the classical
  equations of motion are satisfied, then one has
  $M_{\mu\nu}D_{\nu}=0$, and the Green function (8) may be rewritten
  as
  \be
  G_{\mu\nu}=(\delta_{\mu\lambda}+(\xi-1)D_{\mu}\frac{1}{D^2}D_{\lambda})
  (D^2-2gF)^{-1}_{\lambda\nu}.
  \ee
  The choice $\xi=0$ corresponds to the Landau gauge, in which the
  Green function (9) contains explicitly the projector
  $P_{\mu\lambda}$ onto transverse states:
  \be
  P_{\mu\lambda}=\delta_{\mu\lambda}-D_{\mu}(\frac{1}{D^2})D_{\lambda}.
  \ee

  To define the effective action for the $q\bar q g$ system we use
  the Feynman--Schwinger representation [16]. To do it in proper way
  one should take into account spin degrees of freedom of quarks and
  gluon. Here we omit the spin dependence, reducing the problem to
  the scalar one. This simplified version of the model corresponds to
  the neglecting of colour magnetic interaction (the term
  proportional to the $gF$ in (9)), and omitting the projector.
Similarly, spin dependence in the quark Green function is also
omitted, and we assume
\be
G_q=(D^2-m^2_q)^{-1}.
\ee

As the result, the Feynman--Schwinger representation for the hybrid
Green function takes the form

  \be
  G
  (x_qx_{\bar q} x_g,
  y_qy_{\bar q} y_g)=
  \ee
  $$
  \int^{\infty}_0ds
  \int^{\infty}_0d\bar s
  \int^{\infty}_0d\bar S
  \int DzD\bar z DZ exp(-{\cal{K}}_q-{\cal{K}}_{\bar
  q}-{\cal{K}}_g)<{\cal{W}}>_B,
$$
  where
  $$
  {\cal{K}}_q= m^2_qs+\frac{1}{4}\int^s_0\dot z^2(\tau)d \tau,~~
  {\cal{K}}_{\bar q}= m^2_{\bar q}\bar s+\frac{1}{4}\int^{\bar
  s}_0\dot{\bar  z}^2(\tau)d \tau,
  $$
  $$
  {\cal{K}}_{g}=\frac{1}{4}\int^{S}_0\dot{Z}^2(\tau)d \tau,
  $$
  with boundary conditions
  $$
  z(0)=y_q,~~\bar z(0)=y_{\bar q},~~
  Z(0)=y_g,
  $$
$$   z(s)=x_q,~~{\bar z}(\bar s)=x_{\bar q},~~Z(S)=x_g,
   $$
   and ${\cal{W}}$ is the Wilson loop operator
   \be
   {\cal{W}}=(\lambda_a)^{\alpha}_{\beta}(\Phi_{\Gamma
   q}(y_q,x_q))^{\beta}_{\gamma}
   (\lambda_b)^{\gamma}_{\delta}(\Phi_{\Gamma \bar
   q}(x_{\bar q},y_{\bar q}))^{\delta}_{\alpha}
   (\Phi_{\Gamma g
   }(y_g,x_g)_{ab},
   \ee
   which corresponds to the propagation of quark along the path
   $\Gamma_q$, of antiquark along the path $\Gamma_{\bar q}$ and of
   gluon along the path $\Gamma_g$ (see Fig. ); here $a,b$ are the
   colour indices in the adjoint representation. As in the case of
   $q\bar q$ system [17], all the dependence on the background field
   is contained in the Wilson loop operator (13).

   The Wilson loop configuration (13) may be rewritten using the
   relation between ordered exponents along the gluon path $\Gamma_g$ in
   the adjoint and fundamental representations:
   \be
   \frac{1}{2}(\Phi_{\Gamma
   g}(x,y))_{ab}=(\lambda_a)^{\alpha}_{\beta}(\Phi_{\Gamma
   g}(x,y))^{\beta}_{\gamma}(\lambda_b)^{\gamma}_{\delta}
   (\Phi_{\bar{\Gamma} g} ( y,x))^{\delta}_{\alpha},
   \ee
   where the path $\bar{\Gamma} g$ coincides with the path $\Gamma g$
   and is directed oppositely. The result reads
   \be
   {\cal{W}}=\frac{1}{2}SpW_1SpW_2-\frac{1}{2N_c}SpW,
   \ee
   where $W_1,W_2$ and $W$ are the Wilson loops in the fundamental
   representation along the closed  contours
   $C_1=\Gamma_q\bar{\Gamma}_g,~~
   C_2=\Gamma_{\bar q}{\Gamma}_g$ and
   $C=\Gamma_q\bar{\Gamma}_{\bar q}$ shown at the Figure.

   \section{Generalized area law and effective Hamiltonian}

   To average the Wilson loop configuration (15) over the background
   we use the cluster expansion method generalized in [18] to
   consider the average of more than one Wilson loop. For the
   contours $C_1$ and $C_2$ with the average size much larger than
   the gluonic correlation length $T_g$ we arrive to the generalized
   area law
   \be
   <{\cal{W}}>=\frac{N^2_c-1}{2} exp (-\sigma(S_1+S_2)),
   \ee
   where $\sigma$ is the string tension in the fundamental
   representation, and $S_1$ and $S_2$ are the minimal surfaces
   inside the contours $C_1$ and $C_2$. The area law (16) holds for
   all the configurations in the $q\bar q g$ system, apart from the
   special case of the contours $C_1$ and $C_2$ embedded into the
   same plane,
   where, instead of (16), one has
   \be
   <{\cal{W}}>=\frac{N^2_c-1}{2} exp
   (-\sigma(S_1-S_2))-\sigma^{adj}S_2),~~S_1>S_2,
    \ee
   $\sigma^{adj}$ is the string tension in the adjoint representation.
   The regimes (16) and (17) match smoothly each other at the
   distances between the contours $C_1$ and $C_2$ of order of
   correlation length $T_g$.

   If the string tension is defined mainly by the contribution of
   second order correlators, then $\sigma^{adj}/\sigma=9/4$ for the
   SU(3) colour group. On the other hand, the area law for the Wilson
   loop in the adjoint representation was observed on the lattice
   [19] with $\sigma^{adj}/\sigma\approx 2$, and the same result
   holds true in the limit $N_c\to \infty$. Having all this in mind,
   we assume the regime (16) to be valid everywhere in the $q\bar q
   g$ configuration space.

   The four--dimensional dynamics in (12) can be reduced to the
   three--dimensional one following the procedure used in [17].
   Namely, choosing the physical time parametrization
   $$
   z_{\mu}=(\tau,\vec r_q),~~
   \bar z_{\mu}=(\tau,\vec r_{\bar q}),~~
   Z_{\mu}=(\tau,\vec r_{g})
   $$
   and introducing new dynamical variables
   $$
   \mu_1(\tau)=\frac{T}{2s}\dot z_0(\tau),~~
   \mu_2(\tau)=\frac{T}{2\bar s}\dot{\bar z}_0(\tau),~~
   \mu_3(\tau)=\frac{T}{2S}\dot Z_0(\tau),~~
   $$
   with $0\leq \tau\leq T$, we arrive to the three--dimensional
   representation for the Green function:
   \be
   G=\int D\vec r_q
   D\vec r_{\bar q}
   D\vec r_{g}
   D{\mu_1}
   D{\mu_2}
   D{\mu_3}exp (-A),
   \ee
   with the effective action
   \be
   A=\int^T_0 d\tau\{
   \frac{m_q^2}{2\mu_1}+
   \frac{m_{\bar q}^2}{2\mu_2}+
   \frac{\mu_1+\mu_2+\mu_3}{2}+
   \frac{\mu_1\dot r_q^2}{2}+
   \frac{\mu_r\dot r_{\bar q}^2}{2}+
   \frac{\mu_s\dot r_g^2}{2}+
   \ee
   $$
   +\sigma\int^1_0d\beta_1\sqrt{\dot w_1{w'_1}^2-(\dot w_1w'_1)^2}+
   \sigma\int^1_0d\beta_2\sqrt{\dot w_2{w'_2}^2-(\dot w_2w'_2)^2}\},
   $$
   where the surfaces $S_1 $ and $S_2$ are parametrized by the
       coordinates $w_{i\mu},~~\dot w_{i\mu}=\frac{\partial
       w_{i\mu}}{\partial\tau},~~
       w'_{i\mu}=\frac{\partial
       w_{i\mu}}{\partial\beta_i},~~ i=1,2.$  Assuming the
       straight--line ansatz for the minimal surfaces,
       $$
       w_{1\mu}=\beta_1z_{\mu}+(1-\beta_1)Z_{\mu},~~
       w_{2\mu}=\beta_2\bar z_{\mu}+(1-\beta_2)Z_{\mu},~~
       $$
       we write out the effective Lagrangian for the $q\bar q g$
         system as
         \be
L=   \frac{m_q^2}{2\mu_1}+
   \frac{m_{\bar q}^2}{2\mu_2}+
   \frac{\mu_1+\mu_2+\mu_3}{2}+
   \frac{\mu_1\dot r_q^2}{2}  +
   \frac{\mu_2\dot r_{\bar q}^2}{2}+
   \frac{\mu_3\dot r_g^2}{2}+
   \ee
$$
+\sigma\rho_1\int^1_0d\beta_1\sqrt{1+l_1^2}+
\sigma\rho_2\int^1_0d\beta_2\sqrt{1+l_2^2},
$$
$$
\vec l_1=\frac{1}{\rho_1}
\vec{\rho}_1\times(\beta_1\dot
{\vec r}_q+(1-\beta_1)\dot{\vec r}_g),
$$
$$
\vec l_2=\frac{1}{\rho_2}\vec{\rho}_2\times(\beta_2\dot{\vec
r}_{\bar q}+(1-\beta_2)\dot{\vec r}_g),
$$
$$
\vec{\rho}_1=\vec r_q-\vec r_g,~~~
\vec{\rho}_2=\vec r_{\bar q}-\vec r_g.
$$

To obtain the effective Hamiltonian one should define the momenta and
express the velocities in terms of momenta. It cannot be done
explicitly because of presence of square roots in (20), and to deal
with this problem the auxiliary field approach was suggested in [17].
However, it  was shown in [17] that for the low values of relative
orbital momenta the square roots in (20) can be  expanded up to the
second order in the angular velocities $\vec l_i$, the approximation
proved to be accurate enough even for the massless constituents.
Within this approximation the problem is reduced to the potential --
like one, while the terms $\sim l^2_i$ can be taken into account
perturbatively. The corresponding Hamiltonian in the Minkowsky
space--time in the centre--of--mass frame is easily obtained from
(20):
         \be
H_0= \frac{m_q^2}{2\mu_1}+
   \frac{m_{\bar q}^2}{2\mu_2}+
   \frac{\mu_1+\mu_2+\mu_3}{2}+
   \frac{p^2}{2\mu_p}+\frac{Q^2}{2\mu_Q}+
   \sigma\rho_1+\sigma\rho_2,
   \ee
   $$
   \vec{\rho_1}=\vec{\rho}-\frac{\mu_2}{\mu_1+\mu_2}\vec r,~~
   \vec{\rho_2}=-\vec{\rho}-\frac{\mu_1}{\mu_1+\mu_2}\vec r,~~
   $$
   where the Jacobi coordinates
   $$
   \vec r=\vec r_1-\vec r_2,
   ~~   \vec{\rho_2}=\vec{r}_3-\frac{\mu_1\vec
   r_1+\mu_2\bar r_2}{\mu_1+\mu_2},
   $$
   and conjugated momenta $\vec p$ and $\vec Q$ are introduced,
   and $\mu_p$ and $\mu_Q$ are the reduced masses
   $$
   \mu_p=\frac{\mu_1\mu_2}{\mu_1+\mu_2},
   ~~\mu_Q=\frac{\mu_2(\mu_1+\mu_2)}{\mu_1+\mu_2+\mu_3}.
   $$

   The Hamiltonian still contains the fields $\mu_i(\tau)$, and
   the integration over $\{\mu_i\}$ is to be performed in the
   path integral representation (18) (or, equivalently, taking
   the extremal values of $\mu_i$ in the Hamiltonian). Only after
   that the quantization should be carried out.

   Technically, it is more convenient to proceed in a way
   suggested in [20]: first find the eigenvalues of the
   Hamiltonian (21) assuming $\mu_i$ to be $c$--numbers, and
   after  that minimize the eigenenergies in $\mu_i$. This
   procedure works with rather good accuracy for the lowest
   states, and reduces the problem to the nonrelativistic
   three--body one, with $\mu_i$ playing the role of constituent
   masses. We note that although the Hamiltonian (21) looks like
   the Hamiltonian of the nonrelativistic potential model, it is
   essentially relativistic, and the masses $\mu_i$ are not
   introduced by hand, but are calculated and expressed in terms
   of string tension $\sigma$ and quark masses.

   Another advantage of the above--described method is that it
   allows for the approximate solution to the problem of
   separating out the physical transverse states. Indeed, let us
   impose the constraint
   \be
   \mu_3\Psi_0-\mu_3(\dot{\bar{r}}_g\vec{\Psi})=0
   \ee
   to project out  the physical hybrid state
   $\Psi_{\lambda}=(\Psi_0,\vec{\Psi}),$ where $\lambda$ is the
   gluon spin index. The constraint (22) is compatible with the
   projector $P_{\mu\lambda}$ (10) after averaging over
   background and introducing
   the variables $\mu_i$. In the potential--like regime one has $\vec
   p_g=\mu_3\dot{\bar r}_g$, and we choose the physical states to be
   transverse with respect to the three--dimensional gluon momentum:
   \be
   \vec p_3\vec{\Psi}=0,~~\Psi_0=0.
   \ee
    We are forced to impose the condition (23), because we
   have neglected the spin dependence in the gluon Green
   function (9), so the condition (23) should be treated as
   variational ansatz. The rigorous analysis of the transverse
   and longitudinal gluonic degrees of freedom should be done
   with the inclusion of spin into the path integral
   representation for the gluon Green function.

   \section{Numerical results and discussion}

   In the actual calculations the Hamiltonian was supplied
   with the short range  Coulomb interaction
   \be
   V_c=\frac{\alpha_s}{6r}-
   \frac{3\alpha_s}{2\rho_1}-\frac{3\alpha_s}{2\rho_2}.
   \ee

   As a variational ansatz the Gaussian type radial wave
   functions were chosen. The constraint (23) was satisfied by
   taking the orbital wave functions diagonal in the total
   angular momentum $j$ in the gluonic subsystem, so that the
   states  contain electric or magnetic gluon:
   \be
   \vec{\Psi}^e_j\sim\vec Y_{jjm}(\hat{Q}),~~
   \vec{\Psi}^m_j\sim\sqrt{\frac{j+1}{2j+1}}\vec
   Y_{jj-1m}(\hat Q)+\sqrt{\frac{j}{2j+1}}\vec Y_{jj+1m}(\hat
   Q).
   \ee
   With this choice the electric and magnetic hybrids are
   degenerate, and this degeneracy will be  removed by string
   corrections and by spin--dependent force.

   The quantum numbers of a one--gluon hybrid are given by
   \be
   P=(-1)^{l+j},~~
   C=(-1)^{l+s+1}
   \ee
   for the states with electric gluon, and
   \be
   P=(-1)^{l+j+1},~~
   C=(-1)^{l+s+1}
   \ee
   for the states with magnetic gluon, where $l$ and $s$ are
   the angular momentum and total spin in the
   quark--antiquark subsystem. So the  possible quantum
   numbers for the ground state are
   \be
   J^{PC}=0^{\mp +},1^{\mp +},
   2^{\mp +},1^{\mp -},
   \ee
   where the upper/lower sign stands for the state with
   electric/magnetic gluon (26)/(27).

   The most complicated problem in the constituent approaches is not
   the relative arrangement of ground and excited states, but the
   absolute scale of masses. In the potential model large negative
   constant is needed to fit the $q\bar q$ spectrum, and this
   constant is different for the sectors with different flavour
   content. In the described approach the perimeter terms for the
   Wilson loop and/or  hadronic shifts might be responsible for the
   constant term. We use the prescription that for the hybrid state
   with two strings the additive constant is twice as large as for
   the  $q\bar q$ meson with only one string.

   So, the procedure used is: 1) to define the constant term for the
   given values of parameters from the fit to the $S-,~ P-$ and $D-$
   wave meson levels, and 2) to calculate the hybrid mass with the
   constant multiplied by two.

   The numerical results for the spectra of hybrids with light quarks
   are listed in the Table 1 for different values of quark mass,
   string tension and $\alpha_s$.

   Slightly another procedure was used to calculate the masses of
   hybrids with heavy quarks: the constant term was taken from the
   fit to heavy--light ($D-$ and $B-$) meson masses. We think that
   the "heavy--light" constants are more consistent
   phenomenologically for the hybrid with two "heavy--light" strings.
   The results for the ground states are given in Table 2.

   The spectra we have obtained are rather similar to the ones of the
   flux--tube model [10,11]. There is a lot of common in these two
   approaches, because both models try to account for string
   vibrations. Moreover, the Hamiltonian (21) looks quite similar to
   the Hamiltonian of the "one--bead" flux--tube model [11]. As  it
   was already mentioned, the numerical calculations [11] do not
   support the small oscillation approximation proposed in the
   original version [10] of the flux--tube model. Hence, the
   constraint imposed that the "beads" can oscillate only in  the
   transverse (with respect to quark--antiquark) direction seems to
   be a little suspicious. However, the heavy quark hybrid system was
   analysed [21] in the string--type regime of the Lagrangian (20),
   which matches smoothly at low $j$ the potential--type regime
   described here, and it was shown that the effective values of
   $\sigma\rho_1$ and $\sigma\rho_2$ are equal, and it is just the
   case of the "one--bead" flux--tube model. The difference comes
   from the fact that in the flux--tube the masses of constituents
   (including the bead) are fixed, while in our approach the
   effective masses are the variables. Another numerical discrepancy
   is due to the constant term: in the flux--tube there  is one
   string, so it is reasonable to use the constant fitted by the
   $q\bar q$ spectrum, while we have two distinguishable strings, and
   the constant defined from the $q\bar q$ spectrum should be
   multiplied by the factor of two. These discrepancies compensate
   each other, so the almost exact coincidence of the results seems
   to be to some extent accidental.

   The most important difference between the models is in quantum
   numbers. In the presented picture the confinement is of the
   stochastic nature and is ensured by the background fields, with
   well--defined perturbation theory in the background.
              As a result, the constituent gluon carries quantum
              numbers of its own. On the contrary, the flux--tube,
   being motivated by the strong coupling expansion, knows  nothing
   about the gluons which populate the string. The excitations of the
   string are described by the collective phonon--type modes, so
   instead of (28) one has for the ground state \be J^{PC}= 0^{\mp
   \pm},1^{\mp\pm}, 2^{\mp \pm},1^{\mp\mp}.  \ee The most clear decay
   signature of hybrids is the suppression of a hybrid decay into two
   $S$--wave ground state mesons. This signature takes place for the
   flux--tube hybrid [22] as well as for the electric constituent
   hybrid [23],  and  follows from the symmetry of the wave functions
   involved. It means that $P$--odd hybrids (28) and (29) have the
   same decay properties, and the discrepancy between the models
   should reveal itself in the  $P$-- even  hybrid sector. However,
   all currently discussed hybrid candidates in the light quark
   sector are $P$-- odd!!

   There is growing evidence that hybrids are found at last, the
   belief based mainly on the above--mentioned signature. Indeed, the
   $0^{-+}\pi (1800)$ state seen by VES [24] decays mainly
   into $\pi f_0$  with $\pi\rho$ mode suppressed; the exotic
   $1^{-+}$ signal is seen in BNL [25] in the $\pi f_1$ final state;
   the $\pi(1775) $ is seen in charge exchange photoproduction [26]
   decaying into $\pi f_2$ , that might be $2^{-+}$; the
   $\rho'(1460)$ decays mainly into $\pi a_1$ in contrast to $2~^3S_1
   ~q\bar q$ assignement [27]; the rather promising isoscalar
   $2^{-+}\eta_2(1870)$ [28] is observed with $\pi a_2$ and $\eta
   f_0(980)$ decay models. Unfortunately, up to now  no hybrid--like
   activity is observed in the $P$--even sector.

   \section{Concluding remarks}

   We have demonstrated that the perturbation theory in the
   nonperturbative confining background supports the existence of
   $q\bar q g$ bound states. The hybrid mesonic excitation looks like
   a system of a  gluon with two straight--line strings with quarks
   at the ends. For low values of gluon orbital momentum the problem
   is reduced to the potential--like one, and the resulting hybrid
   spectra are compatible with the data on light quark meson
   spectroscopy.\\

   \begin{center}
   {\bf ACKNOWLEDGEMENTS}
   \end{center}

   We are grateful to Yu.A.Simonov for extremely useful discussions.

   We acknowledge the support from Russian Fundamental Research
   Foundation, Grant No 93-02-14937 and from the International
   Science Foundation and Russian Government, Grant No J77100.
  
 \newpage

   \begin{center}
  {\bf Table 1.} Predicted masses of hybrids with light quarks\\

  \begin{tabular}{|l|l|l|l|l|l|} \hline
  $m_q, GeV$ & $\alpha_s$ & $\sigma, GeV^2$&
  \multicolumn{3}{c|}{$M_{q\bar q g}, GeV$}\\ \cline{4-6}
  &&&$j=1, l=0$&$j=1, l=1$&$ j=2,l=0$\\ \hline
  0&0.3&0.18&1.73&2.03&2.2\\
  &&0.2&1.68&2.00&2.18\\
  0.1&0.3&0.18&1.71&2.02&2.19\\
  &&0.2&1.68&2.00&2.18\\
0  &0.7& 0.18&1.60&1.95&2.15\\
0.1&0.7&0.18&1.58&1.93&2.14
\\ \hline
\end{tabular}

\vspace{2cm}
  {\bf Table 2.} Predicted masses of hybrids with heavy quarks;
  $\sigma=0.18 GeV^2,~~\alpha_s=0.3$\\

  \begin{tabular}{|l|l|} \hline
  $m_c,GeV$&$M_{c\bar c g}, GeV$\\ \hline
  1.2&4.12\\1.5&4.11\\1.7&4.10\\ \hline
  $m_b,GeV$&$M_{b\bar b g}, GeV$\\ \hline
  4.5&10.64\\
  5.2&10.64\\ \hline
  \end{tabular}

  \newpage

  {\bf Figure caption}
  \end{center}

  Wilson loop configuration corresponding to the propagation of the
  hybrid state.
  \end{document}